\documentclass[11pt,twoside]{article}

\usepackage{asp2006}
\usepackage{epsf}

\begin{document}
\title{Measuring H$\beta$ BLR Flows in NGC~5548}
\author{William F. Welsh, Danielle L. Martino\altaffilmark{1}, Go 
Kawaguchi, and}
\affil{San Diego State University, California, USA}
\altaffiltext{1}{Santiago Canyon College, California, USA}
\author{Wolfram Kollatschny}
\affil{Georg-August-Universitat G\"{o}ttingen, Germany}

\begin{abstract}

Using archival {\it{International AGN Watch}} observations we find a 
correlation between the asymmetry of the H$\beta$ line and the system 
luminosity: the line becomes bluer as the system gets brighter. We also 
find a differential lag between the red and blue wings of the line: the 
blue wing lags the red by $\sim$4 days, suggesting an inflow component 
of the BLR.

\end{abstract} 


NGC~5548 is a bright, nearby Sy1 galaxy with a well--known $\sim$10--20~d 
lag between flux variations in the H$\beta$ emission line and the optical 
continuum. 
In this work we re-analyze 5 years of continuum and H$\beta$ fluxes as 
measured and presented in \citet{wfw_Wanders}.
We define the line asymmetry as the blue wing flux / red wing flux,
where the wings span the range ($\pm$) {\mbox{1300--4400~km/s}.}
Asymmetry reversals, corresponding to gross line shape changes, occur 
on a timescale of roughly 2 years, much longer than the light--travel 
time across the BLR. However, the timescale is similar to the dynamical 
timescale suggesting that the asymmetry reversals result from 
large scale flows and changes in structure and/or illumination of the 
BLR, a conclusion previously reached by \citet{wfw_Wanders}.
If the asymmetry is plotted versus the total line flux (see Fig.~1a), 
we find the following correlation: as the line becomes stronger, the line 
becomes bluer.  There is considerable scatter in the trend, some due to 
short timescale reverberation effects, and the trend is only weakly 
present in each year individually. But taken as an ensemble, the 
correlation is clear. Since the line flux is highly correlated with the 
continuum flux, the line becomes bluer as the luminosity increases.
It is important to determine if this correlation is present in other AGN
and in other emission lines.

We computed cross-correlation functions (CCFs) to measure time lags 
between variations in the line and continuum light curves. To patch 
gaps in the light curves we interpolated by using a random walk in flux 
\citep{wfw_Welsh}, motivated by the observation that optical light curves 
of Sy1 AGN exhibit a $\sim 1/f^{2}$ power spectrum \citep{wfw_Collier}. 
We generated 8000 pairs of patched light curves and 
computed standard (local) CCFs. From the distribution of the resulting CCF 
peaks we determined the lag and its uncertainty. This method is inherently 
non-linear and is insensitive to any one particular realization. 
Our lags agree well with previous measurements of the 
H$\beta$ vs.~continuum lag, giving an average lag of 17.2 $\pm$2.6 d.
We then cross correlated the blue wing vs.~the red wing time series and 
found a small, but non-zero, lag (Fig.~1b). As this is a 
{\em{differential}} measurement, it is mostly independent of the continuum 
flux and should be robust. The blue wing lags the red wing by $\sim$4~d
on average, though there is considerable year-to-year scatter. When all 5 
years are analyzed simultaneously the lag is 5.0 $\pm$1.3~d. 
The blue-lags-red signature is highly suggestive of an inflow component 
in the BLR. Inflow signatures have been seen in the CIV line
\citep[e.g.][]{wfw_Koratkar, wfw_Crenshaw, wfw_Korista, wfw_Done} 
but not in the H$\beta$ line. (\citet{wfw_Wanders} {\em{did}} see
a blue-to-red lag of 4.4~d, but considered this to be noise.) 
Note that the hydromagnetic wind {\em{outflow}} model of 
\citet{wfw_Bottorff} can produce a blue-lags-red signature if there is 
extreme anisotropic backscattering such that the near side of the BLR is 
invisible. \citet{wfw_Chiang} presented a model that relies on the 
radiative transfer effects of velocity shear in an optically thick disk 
wind that can also produce a blue-lags-red signature. It would be 
worthwhile to quantitatively examine the simple inflow model and these 
disk+wind outflow models in light of the new H$\beta$ 
asymmetry--luminosity correlation.


\begin{figure}
\plottwo{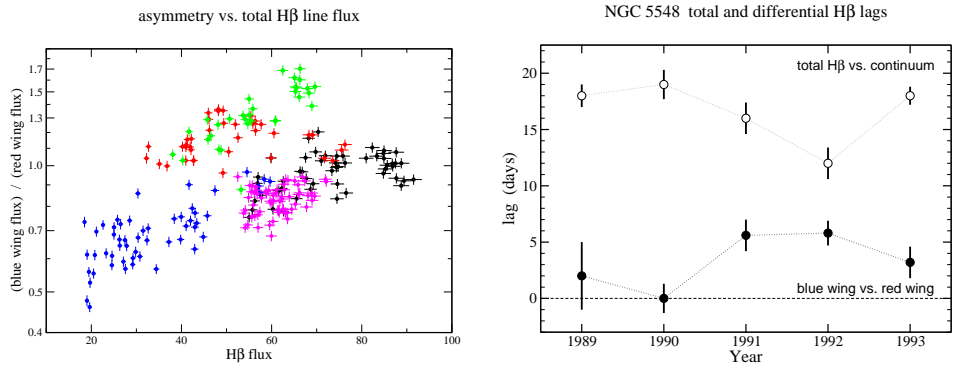}{w_welsh_fig2.eps}
\caption{{\bf Left:} H$\beta$ asymmetry versus total line flux.
{\bf Right:} CCF lags for total H$\beta$ line vs.~continuum flux 
and for blue wing vs.~red wing.
}
\end{figure}


\acknowledgements 
We thank Rob Robinson for useful discussions. This work was supported in 
part by grants AST-0086692 and INT-004905 from the NSF and by grant DFG 
Ko857 from the Deutsche Forschungsgemeinschaft.



\begin{thebibliography}{}

\bibitem[Bentz et al.(2007)]{wfw_Bentz}
Bentz, M. C. et al. 2007, ApJ (in press)

\bibitem[Bottorff et al.(1997)]{wfw_Bottorff}
Bottorff, M., Korista, K. T., Shlosman, I. \& Blandford, R. D. 1997, ApJ, 
497, 200


\bibitem[Chiang \& Murray(1996)]{wfw_Chiang}
Chiang, J. \& Murray, N. 1996 ApJ, 466, 704

\bibitem[Collier \& Peterson(2001)]{wfw_Collier}
Collier, S. \& Peterson, B.M. 2001 ApJ, 555, 775

\bibitem[Crenshaw \& Blackwell(1989)]{wfw_Crenshaw}
Crenshaw, M. \& Blackwell 1989, ApJ, 385, L37

\bibitem[Done \& Krolik(1996)]{wfw_Done}
Done, C. \& Krolik, J. 1996, ApJ, 463, 144

\bibitem[Koratkar \& Gaskell(1989)]{wfw_Koratkar}
Koratkar, A. P. \& Gaskell, C. M. 1989, ApJ, 345, 637

\bibitem[Korista et al.(1995)]{wfw_Korista}
Korista, et al. 1995, ApJS, 97, 285



\bibitem[Wanders \& Peterson(1996)]{wfw_Wanders}
Wanders, I. \& Peterson, B. M. 1996, ApJ, 466, 174

\bibitem[Welsh(2001)]{wfw_Welsh}
Welsh, W. F. 2001 in ``Probing the Physics of Active Galactic Nuclei'', 
ASP Conf. Ser. 224, eds. B. M. Peterson, R. W. Pogge and R. S. Polidan


\end{thebibliography}
\end{document}